\documentclass[twocolumn,showpacs,preprintnumbers,amsmath,amssymb]{revtex4}

\usepackage{graphicx}
\usepackage{dcolumn}
\usepackage{bm}

\begin{document}

\title{Fabrication of graphene nanoribbon by local anodic oxidation lithography using atomic force microscope}

\author{S. Masubuchi,$^1$\footnote{Electronic address: msatoru@iis.u-tokyo.ac.jp} M. Ono,$^1$ K. Yoshida,$^1$ K. Hirakawa,$^{1,2}$ and T. Machida$^{1,2}$\footnote{Electronic address: tmachida@iis.u-tokyo.ac.jp}}
\affiliation{$^1$Institute of Industrial Science, University of Tokyo, 4-6-1 Komaba, Meguro-ku, Tokyo 153-8505, Japan}
\affiliation{$^2$Institute for Nano Quantum Information Electronics, University of Tokyo, 4-6-1 Komaba, Meguro-ku, Tokyo 153-8505, Japan}

\begin{abstract}
We conducted local anodic oxidation (LAO) lithography in single-layer, bilayer, and multilayer graphene using tapping-mode atomic force microscope. The width of insulating oxidized area depends systematically on the number of graphene layers. An 800-nm-wide bar-shaped device fabricated in single-layer graphene exhibits the half-integer quantum Hall effect. We also fabricated a 55-nm-wide graphene nanoribbon (GNR). The conductance of the GNR at the charge neutrality point was suppressed at low temperature, which suggests the opening of an energy gap due to lateral confinement of charge carriers. These results show that LAO lithography is an effective technique for the fabrication of graphene nanodevices.
\end{abstract}

\maketitle

Graphene, a single atomic layer of graphite, has a unique band structure \cite{Novoselov05, Zhang05} and exceptionally high carrier mobility \cite{Bolotin08}. Therefore, it has been used to develop carbon-based electronic devices \cite{Geim07}. To date, graphene nanodevices such as quantum dots \cite{Ponomarenko08, Stampfer08}, Aharonov-Bohm rings \cite{Russo08}, and nanoribbons \cite{Han07, Ling0801} have been fabricated by conventional electron-beam lithography combined with plasma etching. However, plasma etching introduces defects in graphene \cite{Moriki07, Bolotin08, Russo08}, which causes localization of charge carriers \cite{Russo08, Moriki07}. Further, this technique cannot be used to control the edge structure of graphene, which is expected to have significant effects on its electronic properties \cite{Han07}. Therefore, in order to fabricate high-quality devices, we need a new lithography technique that will allow us to perform high-resolution patterning without damaging the graphene layer. 

Local anodic oxidation (LAO) lithography using atomic force microscope (AFM) is a promising technique for the fabrication of graphene nanodevices. This is because LAO lithography has been successfully used for fabricating nanodevices based on semiconductors \cite{Keyser00, Luscher99, Schumacher99, Held97, Fuhrer01}. The confinement of charge carriers obtained by LAO is highly specular \cite{Held99} than that obtained by plasma etching \cite{Thornton89}, {\textit i.e.} the charge carriers conserve their momentum along the normal to the confinement. Recently, Weng {\it et al.} produced insulating trenches in graphene flakes with a thickness of 1-2 atomic layers using tapping-mode AFM \cite{Weng08}. However, in their experiment, the number of graphene layers was not determined, though the LAO conditions such as the width of oxidized area are expected to be totally different for single-layer, bilayer, and multilayer graphene. Geisbers {\it et al.}  used contact-mode AFM to produce an insulating trench on single-layer graphene \cite{Geisbers08}. However, the contact-mode AFM cantilever can damage the fabricated device \cite{Novoselov04}. Further, the transport phenomena of Dirac fermions were not demonstrated clearly in either experiment above \cite{Geisbers08, Weng08}.

In this letter, we describe LAO lithography experiments that were conducted in single-layer, bilayer, and multilayer graphene using tapping-mode AFM. We show that the width of oxidized area depends on the number of graphene layers. We have fabricated an 800-nm-wide bar-shaped device, which exhibits the half-integer quantum Hall effect; this indicates that the conducting channel of graphene is intact during LAO. We have also fabricated a 55-nm-wide graphene nanoribbon (GNR). The conductance of the GNR at the charge neutrality point is strongly suppressed with decreasing temperature from $T$ = 300 to 4.2 K. This suggests the opening of an energy gap due to the lateral confinement of charge carriers. These results show that LAO lithography is an effective technique for the fabrication of graphene nanodevices.

\begin{figure*}[t]
\includegraphics{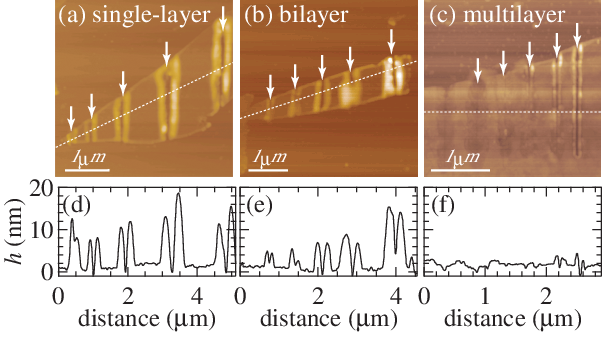}
\caption{(color online) AFM images of (a) single-layer, (b) bilayer, and (c) multilayer graphene. LAO nanolithography was carried out in the direction indicated by the white arrows at the scanning speeds of $v_\textrm{s}$ = 160, 80, 40, 20, and 10 nm/s (left to right). (d), (e), and (f) show the cross-sectional height profiles along the white dashed lines in (a), (b), and (c), respectively.}
\end{figure*}

Graphene flakes were extracted from Kish graphite and deposited onto a 300-nm-thick SiO$_2$ layer by mechanical exfoliation \cite{Novoselov04}. Single-layer, bilayer, and multilayer graphene flakes were identified by the optical color contrast, and the number of graphene layers was verified by measuring the quantum Hall effect \cite{Novoselov05, Zhang05} and Raman spectrum \cite{Ferrari06}. Two-terminal metal electrodes were fabricated by electron-beam lithography (Elionix ELS-7500) followed by thermal evaporation of Au/Cr (40/4 nm). Transport measurements were performed with the standard lock-in technique by applying a small AC current $I_\textrm{AC}$ = 10 nA (18 Hz). A heavily doped Si wafer was used as a global back gate to tune the carrier concentration. Before the transport measurements, surface impurities were removed by annealing the sample at $T$ = 420 K in vacuum ($P$ $\sim$ 1.0 $\times$ 10$^{-4}$ mbar) for several hours.

\begin{figure}[b]
\includegraphics{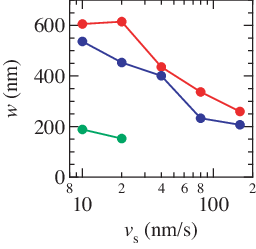}
\caption{(color online) (a) Width of the insulating region $w$ formed in single-layer (red), bilayer (blue), and multilayer (green) graphene as a function of the scanning speed of the AFM cantilever $v_\textrm{s}$.}
\end{figure}

LAO lithography was performed in ambient air by tapping-mode AFM (JEOL JSPM-5200). The relative humidity was maintained at around 39$-$44\%. The spring constant and resonance frequency of the conducting Si cantilever were 42 N/m and 300 kHz, respectively. A positive DC bias voltage (35 V) was applied to the sample during the LAO lithography. The AFM cantilever was scanned with scanning speeds ranging from 10 to 160 nm/s.

Figure 1 shows AFM images of the (a) single-layer, (b) bilayer, and (c) multilayer graphene flakes. LAO lithography was performed by scanning the AFM cantilever in the directions indicated by white arrows at the scanning speeds of $v_\textrm{s}$ = 160, 80, 40, 20, and 10 nm/s (left to right). The cross-sectional height profiles along the white dashed lines in Figs. 1(a)$-$(c) are shown in Figs. 1(d)$-$(f), respectively. When LAO lithography was performed in single-layer graphene, a narrow trench was fabricated, and a bump structure was formed on each side of the trench [Fig. 1(a)]. Trench and bump structures were also produced in bilayer graphene [Fig. 1(b)]. The previous studies on the fabrication of bump structures on HOPG (highly oriented pyrolytic graphite) and graphene \cite{Weng08} did not determine the electronic properties of the bump region. In order to investigate the conductance of the bump regions, we fabricated a device in which carrier transport can occur through a 100-nm-wide and 800-nm-long bump region (not shown). The two-terminal resistance of the device at room temperature was greater than 1 G$\Omega$ independent of the gate bias voltage. This implies that the bump region is an insulating region, which may consist of nonvolatile graphene oxide. Therefore, the effective width of the insulating region $w$ can be obtained by adding the widths of the trench and bump regions.

Figure 2(a) shows $w$ as a function of $v_\textrm{s}$ for single-layer, bilayer, and multilayer graphene. The value of $w$ decreases with $v_\textrm{s}$. This implies that the resolution of LAO lithography increases with $v_\textrm{s}$. Also, $w$ significantly decreases with the number of graphene layers. This suggests that the robustness of graphene flakes against LAO increases with the number of graphene layers.

To investigate the influence of LAO lithography on the electric properties of graphene, we fabricated an 800-nm-wide bar-shaped device as shown in Fig. 3(a). The two-terminal conductance $G$ \cite{Footnote01} decreases sharply at $V_\textrm{g}$ = 5 V [Fig. 3(b)]. This shows that the charge neutrality point of this device is placed close to zero $V_\textrm{Dirac}$ $\sim$ 5 V. The result obtained here is in sharp contrast to that obtained in a device covered with a protective mask of hydrogen silsesquioxane (HSQ) for oxygen plasma etching, where a large positive shift of $V_\textrm{Dirac}$ $>$ 50 V was induced by the presence of HSQ layer even after the annealing process in vacuum \cite{Ling0801}. The electron mobility of the graphene layer is $\mu$ $\sim$ 7000 cm$^2$/Vs at $V_\textrm{g}$ = 25 V, which is comparable to $\mu$ of our single-layer graphene devices measured before LAO lithography. This indicates that the conducting channel of graphene is intact during LAO lithography. Figure 3(c) shows $G$ as a function of $V_\textrm{g}$ measured at $T$ = 4.2 K in a magnetic field of $B$ = 9 T. We can clearly observe the quantum Hall plateaus at $G$ = 2, 6, 10, and 14 $e^2/h$ and $V_\textrm{g}$ = 12, 22, 35, and 45 V, respectively. The quantized plateaus in the sequence $G$ = 4($N$ + 1/2) ($e^2/h$), where $N$ is an integer, are indicative of the half-integer quantum Hall effect. Thus, this result shows that LAO lithography is an efficient technique for the fabrication of graphene devices that exhibit the characteristics of Dirac fermions.

\begin{figure}[t]
\includegraphics{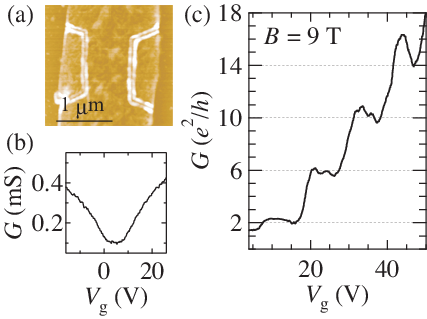}
\caption{(color online) (a) AFM image of the bar-shaped device fabricated in single-layer graphene. (b) Two-terminal conductance $G$ as a function of the gate-bias voltage $V_\textrm{g}$ measured at $T$ = 4.2 K in a zero magnetic field. (c) $G$ as a function of $V_\textrm{g}$ measured at $T$ = 4.2 K and $B$ = 9 T.}
\end{figure}

Finally, we studied the transport properties of graphene nanoribbons (GNRs) whose widths were $W$ = 55 [Fig. 4(b)] and 77 nm (not shown). Each GNR was connected to the metal electrodes through the 2-$\mu$m-wide graphene blocks. Thus, $G$ is restricted by the GNR region. Figure 4(a) shows $G$ of the GNR with $W$ = 55 nm as a function of $V_\textrm{g}$ measured at $T$ = 300 (red), 77 (green), and 4.2 K (blue). At $T$ = 300 K, $G$ is minimum at $G_\textrm{min}$ $\sim$ $4e^2/h$ ($W/L$). When the temperature is decreased to $T$ = 4.2 K, $G_\textrm{min}$ is suppressed by ten times [blue curve in Fig. 4(b)]. The suppression of $G_\textrm{min}$ at low temperature is in sharp contrast to the result that $G_\textrm{min}$ reduces only by 5\% from $T$ = 300 to 4.2 K in the single-layer graphene before LAO (not shown). Figure 4(c) shows the color plot of the differential conductance d$I$$/$d$V$ of the GNR with $W$ = 77 nm as a function of $V_\textrm{g}$ and source-drain bias voltage $V_\textrm{sd}$. In Fig. 4(c), the region of suppressed conductance (dark area) forms diamond-like pattern at around the charge neutrality point. The results presented in Figs. 4(a) and 4(c) quantitatively agree with those observed in GNRs fabricated by oxygen plasma etching \cite{Han07, Molitor08}. Thus, our results suggest the opening of an energy gap $E_\textrm{g}$ due to the lateral confinement of charge carriers. Further, the value of $V_\textrm{sd}$ at the vertex of the dark area gives an energy gap $E_\textrm{g}$ $\sim$ 5 meV [Fig. 4(c)]. This is quantitatively comparable to that obtained in the 60-nm-wide GNR fabricated by the oxygen plasma etching \cite{Han07}. 

\begin{figure}[t]
\includegraphics{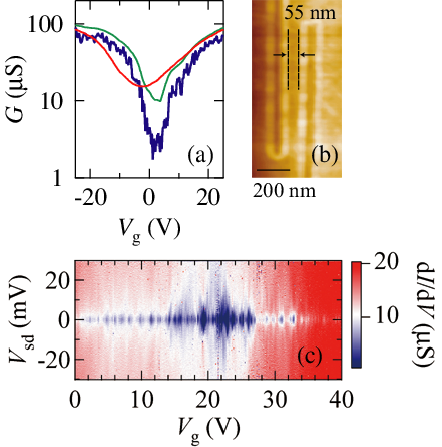}
\caption{(color online) (a) Two-terminal conductance $G$ of the device mentioned in (b) as a function of the gate-bias voltage $V_\textrm{g}$ measured at $T$ = 300 (red), 77 (green), and 4.2 K (blue) in a zero magnetic field. (b) AFM image of a 55-nm-wide ($W$) and 800-nm-long ($L$) nanoribbon formed in single-layer graphene. (c) Color plot of the differential conductance d$I$$/$d$V$ of the 77-nm-wide and 600-nm-long nanoribbon as a function of the gate-bias voltage $V_\textrm{g}$ and the source-drain bias voltage $V_\textrm{sd}$.}
\end{figure}

In summary, we have conducted LAO lithography in single-layer, bilayer, and multilayer graphene using AFM. We showed that the width of the oxidized area depends on the number of graphene layers. We fabricated an 800-nm-wide bar-shaped device, which exhibits the half-integer quantum Hall effect. We also fabricated a 55-nm-wide graphene nanoribbon (GNR). The conductance of the GNR at the charge neutrality point was strongly suppressed by decreasing $T$. These results show that LAO lithography is an effective technique for the fabrication of graphene nanodevices.

The authors acknowledge M. Kawamura for technical support. This study is supported by Grants-in-Aid from MEXT, the Special Coordination Funds for Promoting Science and Technology, and the CREST, Japan Science and Technology Agency. One of the authors (S. M.) acknowledges the JSPS Research Fellowship for Young Scientists.


\begin{thebibliography}{99}
\bibitem{Footnote01} The contribution of the contact resistance $R_\textrm{cont}$ = 1.5 k$\Omega$ to $G$ was compensated as $G$ = $(R_\textrm{2p}$ - $R_\textrm{cont}$)$^{-1}$, where $R_\textrm{2p}$ is the measured two-terminal resistance.
\end{thebibliography}
\end{document}